\documentclass{article}

\PassOptionsToPackage{numbers, compress}{natbib}

\usepackage[dblblindworkshop, final]{neurips_2025}
\workshoptitle{ML4PS}

\usepackage[utf8]{inputenc} 
\usepackage[T1]{fontenc}    
\usepackage{hyperref}       
\usepackage{url}            
\usepackage{booktabs}       
\usepackage{amsfonts}       
\usepackage{nicefrac}       
\usepackage{microtype}      
\usepackage{xcolor}         

\newcommand{\fO}[1]{$\mathcal{O}(N^{#1})$ }
\usepackage{longtable}
\usepackage{subcaption}
\usepackage{siunitx}
\usepackage{tabularx,array}
\newcolumntype{Y}{>{\raggedright\arraybackslash}X} 

\usepackage[percent]{overpic}

\usepackage{graphicx}
\graphicspath{{figures/}{./}}
\title{
$\Delta$-ML Ensembles for Selecting Quantum Chemistry Methods to Compute Intermolecular Interactions
}

\author{%
  Austin M.~Wallace\thanks{Center for Computational Molecular Science and Technology
    and School of Chemistry and Biochemistry} \\
  School of Chemistry and Biochemistry\\
  Georgia Institute of Technology\\
  Atlanta, GA 30332-0400 \\
  \texttt{awallace43@gatech.edu} \\
  \And
  C. David.~Sherrill\thanks{Center for Computational Molecular Science and Technology,
    School of Chemistry and Biochemistry,
    and School of Computational Science and Engineering
} \\
  School of Chemistry and Biochemistry\\
  Georgia Institute of Technology\\
  Atlanta, GA 30332-0400 \\
  \texttt{sherrill@gatech.edu} \\
  \And
  Giri P. ~Krishnan
  \\
  Center for Artificial Intelligence in Science and Engineering\\
  Georgia Institute of Technology\\
  Atlanta, GA 30308 \\
  \texttt{giri@gatech.edu} \\
}

\begin{document}

\maketitle

\begin{abstract}

    \emph{Ab initio} quantum chemical methods for accurately computing
    interactions between molecules have a wide range of applications but are
    often computationally expensive. Hence, selecting an appropriate method
    based on accuracy and computational cost remains a significant challenge
    due to varying performance of methods. In this work, we propose a framework
    based on an ensemble of $\Delta$-ML models trained on features extracted
    from a pre-trained atom-pairwise neural network to predict the error of
    each method relative to all other methods including the ``gold standard''
    coupled cluster with single, double, and perturbative triple excitations at
    the estimated complete basis set limit [CCSD(T)/CBS]. Our proposed approach
    provides error estimates across various levels of theories and identifies the
    computationally efficient approach for a given error range utilizing only a
    subset of the dataset. Further, this approach allows comparison between
    various theories. We demonstrate the effectiveness of our approach using an
    extended BioFragment dataset, which includes the interaction energies for
    common biomolecular fragments and small organic dimers. Our results show
    that the proposed framework achieves very small mean-absolute-errors below
    0.1 kcal/mol regardless of the given method. Furthermore, by analyzing
    all-to-all $\Delta$-ML models for present levels of theory, we identify
    method groupings that align with theoretical hypotheses, providing evidence
    that $\Delta$-ML models can easily learn corrections from any level of
    theory to any other level of theory. 

\end{abstract}


\section{Introduction}

Accurate quantum mechanical (QM) computations of intermolecular interactions
are valuable to identify the most probable crystal structure for organic
molecules,\cite{Hoja:2016:e70057, Borca:2023:234102} understanding
protein-ligand interactions involved in binding,\cite{Meyer:2003:e70057,
Parrish:2017:7887}, modeling nucleotide stacking,\cite{Hill:2003:729,
Parker:2013:1306} and developing intermolecular
force-fields.\cite{McDaniel:2013:2053, VanVleet:2018:739, Schriber:2021:184110}
Although many methods exist to compute interaction energies, the trade-off of
accuracy and computational cost drives the choice of specific pairings of
methods and basis sets for quantum mechanical calculations. Any specific
method/basis set pair is called the level of theory.
CCSD(T)/CBS\cite{Raghavachari:1989} is considered the gold standard level of
theory for interaction energies,\cite{Rezac:2013:2151}; however, it scales as
\fO{7}, making it very expensive.

Within QM, the interaction energy quantifies how attractive or
repulsive two molecules are to each other. More formally, the interaction
energy can be defined in a supermolecular approach through
\begin{equation}
    \Delta E_{\rm int} = E_{IJ} - E_{I} - E_{J},
\end{equation}
where $IJ$ represents the energy of a dimer while $I$ and $J$ represent the
energies of the isolated monomers. The types of non-covalent interactions that
impact the interaction energy are electrostatics, van der Waals forces,
hydrogen bonds, exchange-repulsion---akin to steric energies---and
polarization.

QM interaction energies are quite sensitive to electron correlation, basis set
size, and counterpoise corrections (CP).\cite{Burns:2014:49} Consequently,
predicting interaction energies from lower levels of theory, such as
Hartree-Fock (HF), can lead to significant errors, while sometimes inexpensive
methods relying on error cancellation like SAPT0/jun-cc-pVDZ can yield
reasonably accurate results in certain chemical systems, while failing at
others, like $\pi-\pi$ aromatic systems.\cite{Schriber:2025:084114} For small
systems, high-accuracy methods like CCSD(T)/CBS can be computed; however, the
scaling of \fO{7} makes these methods intractable for most practical
applications. Hence, high-throughput screening approaches largely rely on the
most inexpensive QM methods like HF, MP2, or DFT even at the cost of accuracy.
With hundreds of levels of theory available, selecting an appropriate one for
any particular set of chemical systems becomes a significant challenge,
especially for novice users. In this work, we demonstrate the effectiveness of
$\Delta$-ML neural network models which leverage pre-trained models for QM
interaction energies and predict the difference between lower accuracy method
and higher accuracy method, providing a significant computational gain without
major loss in accuracy. The $\Delta$-ML neural network models can be trained on
a small subset of the data and provide strong generalization, enabling
potential use in large-scale screening of molecules.

\subsection{Key Contributions}

\begin{itemize}
    \item Our framework identifies appropriate levels of theory for a given system through a
        combination of compute time estimators and  $\Delta$-ML error predictions.
    \item Hierarchical clustering of the $\Delta$-ML Ensemble demonstrates
        these models capture theoretical relationships between methods,
        providing evidence for the effectiveness of applying $\Delta$-ML models
        to identify computationally efficient levels of theory for
        chemical system(s).
\end{itemize}


\textbf{Related Works}: Machine-learned $\Delta$-correction models have emerged as a potential approach
to predict the result of accurate methods from less expensive methods using
neural networks or machine learning.\cite{Ramakrishnan:2015:2087,
Nandi:2021:051102, Cheng:2019:131103, Vinod:2025:024134, Huang:2025:035004}
Such $\Delta$-ML methods allow capturing expensive electron
correlation effects\cite{Ramakrishnan:2015:2087} and basis set effects.
Oftentimes only a very small percentage of the dataset is needed to be computed
at the higher level of theory.\cite{Ramakrishnan:2015:2087, Song:2023:11192} In
such methods, the objective is to predict the difference (or $\Delta$) between
the target high-level of theory interaction energy ($E_{\textrm{high}}$) and a
low-level of theory ($E_{\textrm{low}}$) using machine learning methods.
This task assumes that there are computationally inexpensive functions that can capture more expensive functions, such as, high-level electron correlation in terms of molecular features relating to the geometry and pre-training on other properties.


Interaction energies present unique challenges in which approximate levels of
theory can yield overbinding or underbinding due to combinations of incomplete
correlation effects, basis set truncation errors, and types of interactions based 
on the chemical system.\cite{Burns:2014:49, Parker:2014:094106,
Schriber:2025:084114} As a result, naive models trained to predict total
energies do not necessarily yield accurate predictions for interaction
energies. The present $\Delta$-ML models address this issue by focusing
directly on the discrepancies in $E_{\textrm{int}}$, exploiting the smoother
error landscape of the delta compared to the total energy.

The present work targets developing $\Delta$-ML deep neural network models to
predict interaction energies of one level of theory from another. Generally,
$\Delta$-ML models are targeting a single level of theory to a reference level
of theory; however, the present work expands this to 80 levels of
theory to acquire additional insight into how levels of theory
compare for interaction energies. Typically, one would want to predict the
expected error from a lower-level of theory to a higher-level of theory,
but one could also ask how well can one map from any level of theory to
another. The models do not require interaction energies as inputs to
compute the error; hence, an additional application of these models is to
estimate how inaccurate a level of theory would be if computed prior to any
quantum calculations.

\section{Methods}


\textbf{Dataset}: The present work leverages data accumulated through various different works\cite{Thanthiriwatte:2011:88,
    Marshall:2011:194102, Marshall:2011:194102, Burns:2011:084107,
    Smith:2016:2197, Jurecka:2006:1985, Grafova:2010:2365, Smith:2016:2197,
Burns:2017:161727}  
to investigate how 80 different levels of theory perform at predicting
intermolecular interaction energies on small organic molecules. More specifics
on the subsets are available in Table S1. The dataset contains 3816 dimers with
reference data at approximately ``silver standard'' interaction energies
[DW-CCSD(T**)-F12/aug-cc-pVDZ]. However, to acquire the gold standard energies,
the present work computed a subset of 3324 dimers with CCSD(T)/CBS/CP for
higher quality reference energies. Methods are paired with specific Dunning's
augmented, correlation consistent double, triple, or quadruple-$\zeta$ basis
sets\cite{Dunning:1989, Woon:1994:2975}---cc-pVDZ, aug-cc-pVDZ, aug-cc-pVTZ,
and aug-cc-pVQZ. From herein, the present work will refer to this dataset as
BFDB-Ext, containing 250K quantum interaction energy computations made easily
accessible through this work. Due to the dataset on small organic dimers up to
38 atoms consisting of H, C, N, O, and S, the developed models are not
guaranteed to generalize to significantly larger molecular systems like
biomolecules.

\textbf{$\Delta$-corrected Models Ensembles from Pre-trained Models}: To
provide a reliable recommendation of an appropriate level of theory for
computing intermolecular interaction energies, it is necessary to estimate the
errors associated with each method relative to established reference values
based on experimental measurements or computational benchmarks. Using BFDB-Ext,
models can be trained to estimate the error for a given dimer using a
particular level of theory where $E_{\rm IE, ref}$ is CCSD(T)/CBS/CP. For each
level of theory, a separate $\Delta$-model is trained to predict the error
through
\begin{equation}
    \Delta E_{\rm pred} \approx E_{\rm IE, x} - E_{\rm IE, ref},
    \label{eq:error}
\end{equation}
where $E_{\rm IE, x}$ is the interaction energy at the specified level of
theory.


We employ a pre-trained model originally developed for predicting dimer
interaction energies on a substantially larger and more diverse dataset. This
allows for our framework to be applicable to smaller datasets which may have limited
chemical diversity.
A recent atomic-pairwise neural network (AP-Net2) model is a 2.6M parameter
pre-trained model that employs message-passing networks to predict monomer properties and subsequently
SAPT0/jun-cc-pVDZ interaction energies.\cite{Glick:2024:13313} AP-Net2 was
trained the Splinter dataset\cite{Spronk:2023:619} of over 1.6 million
datapoints from over 9000 unique dimers, primarily targeting describing
protein-ligand interactions. 
Since BFDB-Ext molecules resemble those in the Splinter dataset, AP-Net2
embeddings are well suited for $\Delta$-corrected model for BFDB-Ext dataset.


Hyperparameter search identified a five-layer network (details in Supplement)
as sufficient to achieve errors below 0.1 kcal/mol. Models were trained for 100
epochs on a 40/60 train/test split using mean squared error (MSE) between
levels of theory as the loss, with inputs taken from the penultimate embeddings
of AP-Net2. To train all-to-all $\Delta$-ML models requires approximately 450
walltime hours with 8 cores on a Xeon 6226 CPU. In future works, the total
number of levels of theory for larger datasets would be limited based on some
methods having similar error distributions and allowing the approach to 
generalize to more data.

\begin{figure*}
    \includegraphics[width=0.99\textwidth]{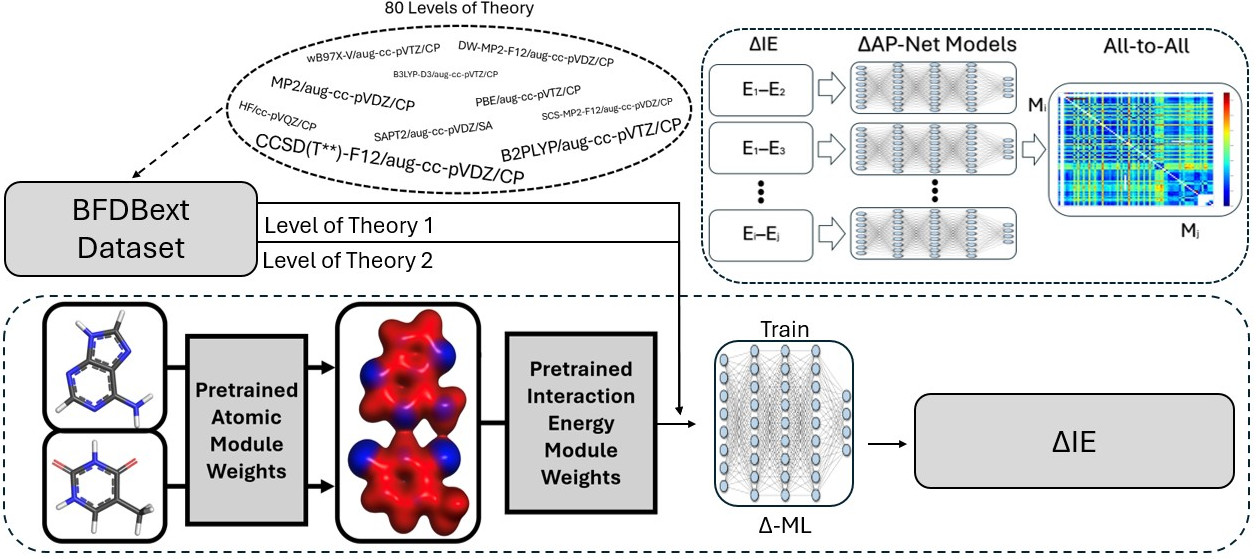}
    \caption{
        Overview of methodology of using the BFDBext to train 80x80
        $\Delta$AP-Net2 models for predicting from any level of theory in
        the dataset to another level of theory.
    }
    \label{fg:overview}
\end{figure*}


\textbf{Compute Time Estimators}: 
Alongside error estimation, we fit a polynomial to compute times using water
clusters and small organics from BFDB-Ext. 
This task is necessary for downstream applications of
the error estimating model by restricting recommended levels of theory to those
that are computable by the end user. Otherwise, the error estimator would
always recommend using CCSD(T)/CBS/CP energies, although in reality this is
not desirable nor realistically computable for many chemical systems.

\section{Results \& Discussion}

\begin{figure*}[t!]
    \centering
    \begin{subfigure}[t]{0.49\textwidth}
\phantomsubcaption\label{fg:test} 
    \begin{overpic}[width=\linewidth]{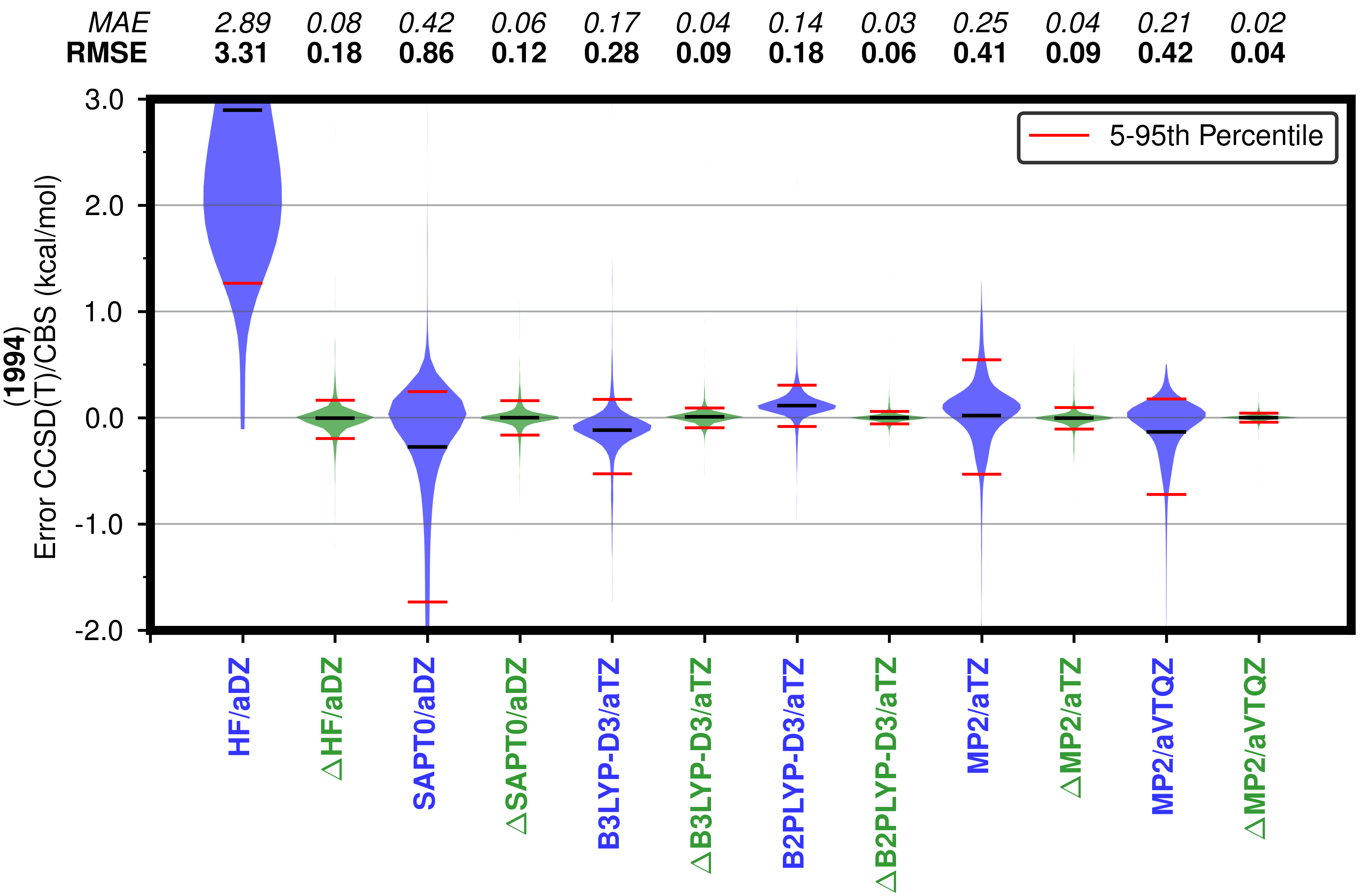}
      \put(2,0){\small\bfseries(\alph{subfigure})}
    \end{overpic}
    \end{subfigure}\hfill
    \begin{subfigure}[t]{0.49\textwidth}
    \phantomsubcaption\label{fg:dendogram}
    \begin{overpic}[width=\linewidth]{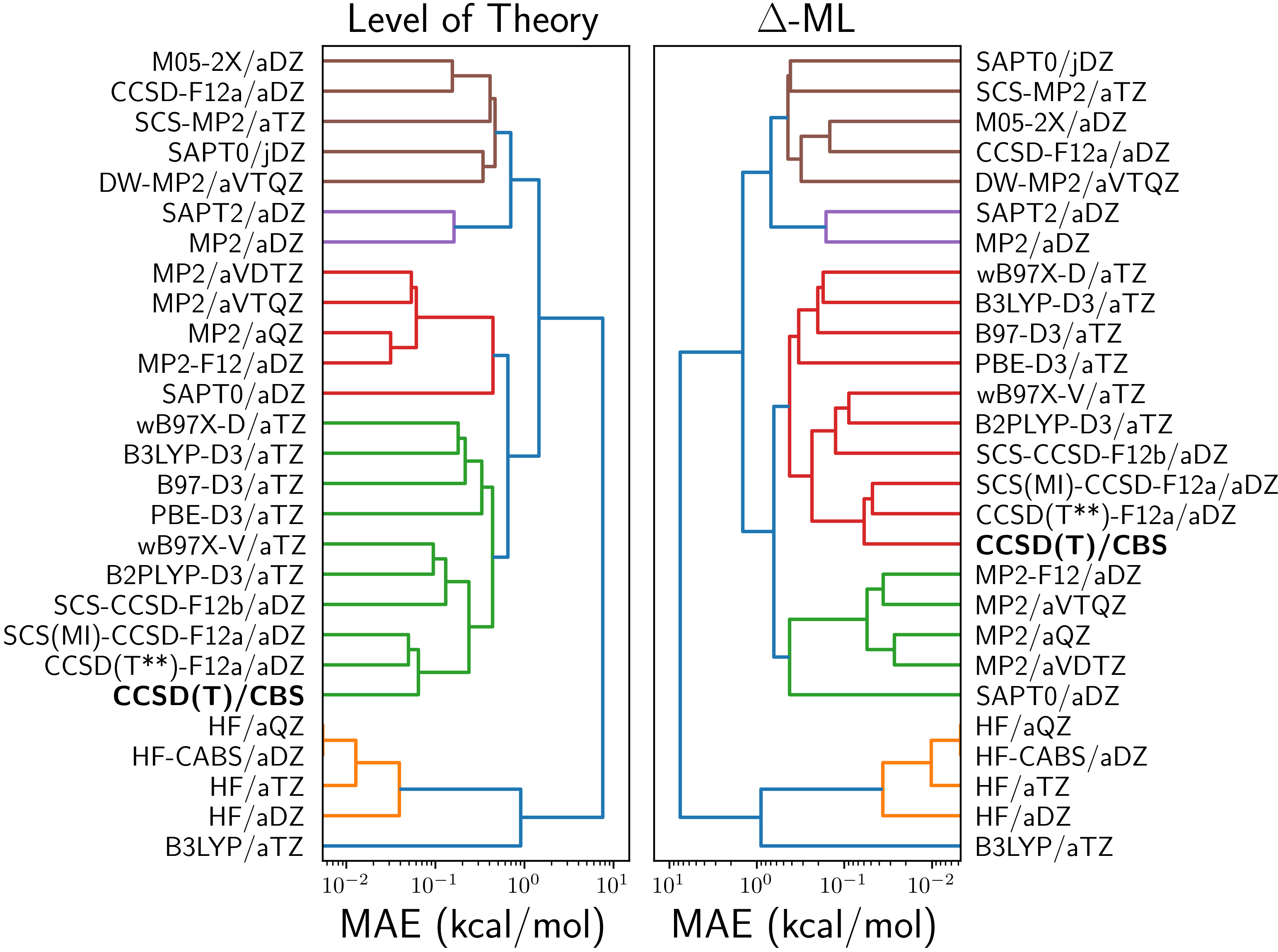}
      \put(2,9){\small\bfseries(\alph{subfigure})}
    \end{overpic}
    \end{subfigure}
    \caption{
        (a) BFDBExt dataset test error distributions for select levels of theory
        with respect to an estimated CCSD(T)/CBS/CP reference. The black
        horizontal line represents the mean error and the red horizontal lines
        represent the 5th and 95th percentiles. The uncorrected level of theory
        IE errors are in blue, while the $\Delta$AP-Net2 plus level of theory IE
        errors are in green.
        (b) Dendogram of select methods $\Delta$AP-Net2 model predicted error
        estimations ordered by MAE. Note the clusters of methods are
        nearly identical as the all-to-all M1 to M2 dendogram in the SI,
        meaning that the models are accurately predicting any M1 to M2. All levels of theory here are using CP.
}
\end{figure*}

\textbf{Model Performance}: Selected $\Delta$AP-Net2 models are shown in Figure
\ref{fg:test} demonstrating performance predicting electron correlation
corrections from a base level of theory to the reference, which are estimated
CCSD(T)/CBS/CP energies in this case. Particularly different classes of
methods---HF, MP2, SAPT, B3LYP, and B2PLYP---are included in the primary table
(full list included in Table S2). Even HF/aug-cc-pVDZ/CP can be corrected from
an MAE of 2.89 kcal mol$^{-1}$ to 0.08 kcal mol$^{-1}$, albeit still having a
max error of 4.09 kcal mol$^{-1}$. Meanwhile, other levels of theory that have
better baseline errors can also be corrected to roughly the same accuracy, but
smaller max errors. For example, MP2/aug-cc-pVQZ/CP  has a baseline MAE of 0.21
kcal mol$^{-1}$ and a max unsigned error of 3.56 kcal mol$^{-1}$, but after
applying a $\Delta$AP-Net2 model, the MAE is reduced to 0.02 and max error to
0.73 kcal mol$^{-1}$. The models accurately predict errors (below <0.1 Kcal) on
the test set, effectively learning the mapping from one level of theory to the
reference. Here we tested the generalization only within the same  chemical
spaces and further work could extend this framework to evaluate generalization
to disparate chemical spaces where the mapping might be more complex. 

\textbf{Level of Theory Hierarchies}: 
Clustering of the MAE from all-to-all predictions, we evaluated how well the
$\Delta$-ML ensemble captures the relationships between different levels of
theory compared to theoretical expectations. As shown in Figure
\ref{fg:dendogram}, the dendrograms from both the $\Delta$-ML and theoretical
expectation show strong alignment. This shows that the $\Delta$-ML models
capture relationships between levels of theory, further validating the approach
(see SI for details).


\textbf{Time Estimation}: While predicting the exact compute time for a given level of theory would
require detailed knowledge of the hardware and software implementation, a rough estimate
can be acquired by fitting polynomials to accurately predict the log of the 
compute times. The practical goal of this task is to filter out levels of
theory that are beyond the user's computational budget. To this end, polynomial
expressions detailed in the Appendix are of the
available singlepoint energy computations on water clusters and small organic
molecules from the BFDBext dataset. The resulting fitting RMSEs are shown in
Table S3. While the fits are not perfect, they reasonably filter out
levels of theory that are too expensive for given systems.

\section{Conclusion}

The present work has demonstrated that $\Delta$-ML models can be trained to
predict the error of a given level of theory from any other level of theory.
Particularly, the models are able to use one of the cheapest levels of theory,
HF/aug-cc-pVDZ/CP, to predict CCSD(T)/CBS/CP reference value with a
surprisingly small MAE of 0.08 kcal mol$^{-1}$. Even more interesting is that
these models are able to predict between any two levels of theory with similar
accuracy even when the methods themselves quite differently like DFT to
wavefunction methods on these systems. Furthermore, when combining the ensemble
of $\Delta$-ML models with the compute time estimators, users can rely on data
instead of chemical intuition to select an appropriate level of theory for
their desired accuracy, computational cost, and chemical system(s). To enhance
generalization, this framework can be applied to datasets with more chemical
diversity and likely fewer levels of theory.  A next step of this work is to
unify the usage of error and time estimators to enable large-scale screening
applications critical for material or drug discovery.

\bibliographystyle{achemso}
\bibliography{refs}


\appendix

\section{Technical Appendices and Supplementary Material}

\subsection{Model Details}

The $\Delta$-ML models used within this work are based on the atom-pairwise
message passing neural networks developed in previous
work.\cite{Glick:2024:13313} These consist of an atomic module that learns to
predict atomic charges, dipoles and quadruples through message-passing neural
networks. This module uses 3 message passes, 8 Bessel functions, and a cutoff
distance of 5.0 \AA. The update and readout functions are dense feed-forward
neural networks with 3 three hidden layers with 256, 128, and 64 neurons. The
last layer has a linear operation to reach the last hidden layer of size 8 or 1
for update and readout, respectively. The intermolecular atomic-pairwise module
that has been adapted for the $\Delta$-ML models use the same defaults as
AP-Net2, except for predicting a single energy instead of 4 and dropping the
multipolar electrostatics. The $\Delta$-ML update and readout layers use the
same hidden layer sizes as the atomic module.

\begingroup
\begin{table*}[ht!]
    \centering \footnotesize
    \caption{Datasets used in training $\Delta$-ML models. For each dataset,
        we provide the total number of dimers (Size), the number of heavy
        atoms in the largest dimer (Largest), relevant references, and a brief
        description.}
    \begin{tabularx}{\textwidth}{l c c l Y}
        \hline\hline
        Database & Size & Largest & Ref. & Description \\ \hline
        \multicolumn{1}{l}{\em Curves \& Surfaces}\\
        HBC6 & 118 & 6  & \cite{Thanthiriwatte:2011:88, Marshall:2011:194102} & dissoc. curves of doubly hydrogen-bonded (HB) complexes \\
        NBC10ext & 183 & 12 & \cite{Marshall:2011:194102, Burns:2011:084107, Smith:2016:2197} & dissoc. curves of dispersion-bound (DD) complexes \\
        \multicolumn{1}{l}{\em Small Dimers}\\
        S22 & 22 & - & \cite{Jurecka:2006:1985, Grafova:2010:2365, Smith:2016:2197} & \\
        \multicolumn{1}{l}{\em Extracted from Biological Systems}\\
        SSI & 3372 & 20 & \cite{Burns:2017:161727} & peptide sidechain-sidechain complexes \\
        BBI & 100 & 20 & \cite{Burns:2017:161727} & peptide sidechain-sidechain complexes \\
        \multicolumn{1}{l}{\em Total} & 3816 & 20 & & \\
        \hline\hline
    \end{tabularx}
    \label{tab:db}
\end{table*}
\endgroup

\begin{longtable}{|c|c|c|}
\hline
\textbf{Method} & \textbf{Basis Set} & \textbf{Mode} \\
\hline
B2PLYP-D3 & aug-cc-pVTZ & CP \\
\hline
DW-CCSD(T**)-F12 & aug-cc-pVDZ & CP \\
\hline
CCSD(T**)-F12a & aug-cc-pVDZ & CP \\
\hline
MP2 & aug-cc-pVTQZ & CP \\
\hline
CCSD-F12a & aug-cc-pVDZ & CP \\
\hline
HF-CABS & aug-cc-pVDZ & CP \\
\hline
SCS(MI)-MP2 & cc-pVQZ & CP \\
\hline
DW-MP2 & cc-pVQZ & CP \\
\hline
SCS(N)-MP2 & cc-pVQZ & CP \\
\hline
SCS-MP2 & cc-pVQZ & CP \\
\hline
HF & cc-pVQZ & CP \\
\hline
MP2 & cc-pVQZ & CP \\
\hline
SCS(MI)-MP2 & aug-cc-pVTZ & CP \\
\hline
DW-MP2 & aug-cc-pVTZ & CP \\
\hline
SCS(N)-MP2 & aug-cc-pVTZ & CP \\
\hline
SCS-MP2 & aug-cc-pVTZ & CP \\
\hline
HF & aug-cc-pVTZ & CP \\
\hline
MP2 & aug-cc-pVTZ & CP \\
\hline
SCS(MI)-CCSD-F12a & aug-cc-pVDZ & CP \\
\hline
SCS(MI)-CCSD-F12b & aug-cc-pVDZ & CP \\
\hline
DW-MP2 & aug-cc-pVDZ & CP \\
\hline
SCS-CCSD-F12b & aug-cc-pVDZ & CP \\
\hline
MP2-F12 & aug-cc-pVDZ & CP \\
\hline
CCSD-F12b & aug-cc-pVDZ & CP \\
\hline
SCS(N)-MP2 & aug-cc-pVDZ & CP \\
\hline
CCSD(T**)-F12b & aug-cc-pVDZ & CP \\
\hline
SCS-MP2-F12 & aug-cc-pVDZ & CP \\
\hline
SCS-MP2 & aug-cc-pVDZ & CP \\
\hline
DW-MP2-F12 & aug-cc-pVDZ & CP \\
\hline
SCS-CCSD-F12a & aug-cc-pVDZ & CP \\
\hline
HF & aug-cc-pVDZ & CP \\
\hline
MP2 & aug-cc-pVDZ & CP \\
\hline
SCS(N)-MP2-F12 & aug-cc-pVDZ & CP \\
\hline
SCS(MI)-MP2 & aug-cc-pVDTZ & CP \\
\hline
DW-MP2 & aug-cc-pVDTZ & CP \\
\hline
SCS(N)-MP2 & aug-cc-pVDTZ & CP \\
\hline
SCS-MP2 & aug-cc-pVDTZ & CP \\
\hline
MP2 & aug-cc-pVDTZ & CP \\
\hline
SCS(MI)-MP2 & aug-cc-pVQZ & CP \\
\hline
DW-MP2 & aug-cc-pVQZ & CP \\
\hline
SCS(N)-MP2 & aug-cc-pVQZ & CP \\
\hline
SCS-MP2 & aug-cc-pVQZ & CP \\
\hline
HF & aug-cc-pVQZ & CP \\
\hline
MP2 & aug-cc-pVQZ & CP \\
\hline
SCS(MI)-MP2 & aug-cc-pVTQZ & CP \\
\hline
DW-MP2 & aug-cc-pVTQZ & CP \\
\hline
SCS(N)-MP2 & aug-cc-pVTQZ & CP \\
\hline
SCS-MP2 & aug-cc-pVTQZ & CP \\
\hline
SAPT0 & aug-cc-pVDZ & SA \\
\hline
SAPT0 & jun-cc-pVDZ & SA \\
\hline
sSAPT0 & aug-cc-pVDZ & SA \\
\hline
sSAPT0 & jun-cc-pVDZ & SA \\
\hline
SCS-SAPT0 & jun-cc-pVDZ & SA \\
\hline
SAPT2 & aug-cc-pVDZ & SA \\
\hline
SAPT2+ & aug-cc-pVDZ & SA \\
\hline
B3LYP & aug-cc-pVTZ & unCP \\
\hline
B3LYP-D2 & aug-cc-pVTZ & unCP \\
\hline
B3LYP-D3 & aug-cc-pVTZ & unCP \\
\hline
B2PLYP & aug-cc-pVTZ & unCP \\
\hline
B2PLYP-D2 & aug-cc-pVTZ & unCP \\
\hline
B2PLYP-D3 & aug-cc-pVTZ & unCP \\
\hline
B97 & aug-cc-pVTZ & unCP \\
\hline
wB97X-D & aug-cc-pVTZ & unCP \\
\hline
M05-2X & aug-cc-pVDZ & unCP \\
\hline
PBE & aug-cc-pVTZ & unCP \\
\hline
PBE-D2 & aug-cc-pVTZ & unCP \\
\hline
PBE-D3 & aug-cc-pVTZ & unCP \\
\hline
B97-D2 & aug-cc-pVTZ & unCP \\
\hline
B97-D3 & aug-cc-pVTZ & unCP \\
\hline
B2PLYP & aug-cc-pVTZ & CP \\
\hline
B3LYP & aug-cc-pVTZ & CP \\
\hline
B3LYP-D3 & aug-cc-pVTZ & CP \\
\hline
B97-D3 & aug-cc-pVTZ & CP \\
\hline
M05-2X & aug-cc-pVDZ & CP \\
\hline
PBE & aug-cc-pVTZ & CP \\
\hline
PBE-D3 & aug-cc-pVTZ & CP \\
\hline
wB97X-D & aug-cc-pVTZ & CP \\
\hline
wB97X-V & aug-cc-pVTZ & CP \\
\hline
wB97X-V & aug-cc-pVTZ & unCP \\
\hline
CCSD(T) & CBS & CP \\
\hline
\caption{List of all levels of theory, basis sets, and modes used in the}
\end{longtable}

\begin{table}[h!]
\centering
\begin{tabular}{|l|c|c|}
\hline
\textbf{Level of Theory} & \textbf{Train RMSE [log(s)]} & \textbf{Test RMSE [log(s)]} \\
\hline
MP2 & 0.1542 & 0.1855 \\
\hline
HF & 0.1048 & 0.1175 \\
\hline
B2PLYP-D3 & 0.1518 & 0.1444 \\
\hline
B3LYP-D3 & 0.1966 & 0.1875 \\
\hline
PBE-D3 & 0.2005 & 0.1817 \\
\hline
M05-2X & 0.2148 & 0.2021 \\
\hline
wB97X-V & 0.2025 & 0.1851 \\
\hline
wB97X-D & 0.1812 & 0.1531 \\
\hline
FNO-CCSD & 0.1811 & 0.1687 \\
\hline
FNO-CCSD(T) & 0.2404 & 0.1916 \\
\hline
\end{tabular}
\caption{Summary of polynomial fitting errors for different levels of theory}
\end{table}

\begin{figure*}[h]
    \includegraphics[width=0.99\textwidth]{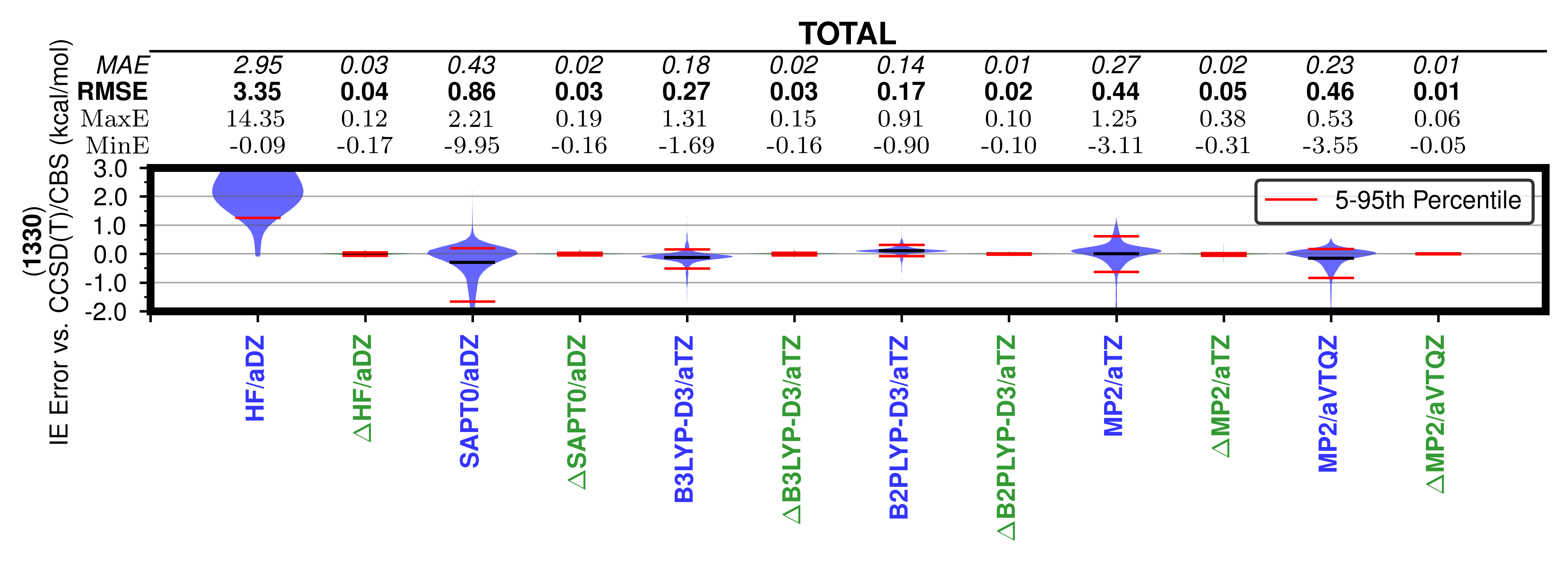}
    \caption{
        BFDBExt dataset train error distributions for select levels of theory
        with respect to an estimated CCSD(T)/CBS/CP reference. The black
        horizontal line represents the mean error and the red horizontal lines
        represent the 5th and 95th percentiles. The uncorrected level of theory
        IE errors are in blue, while the $\delta$AP-Net2 plus level of theory IE
        errors are in green.
    }
    \label{fg:train}
\end{figure*}

\begin{figure*}[h]
    \includegraphics[width=0.99\textwidth]{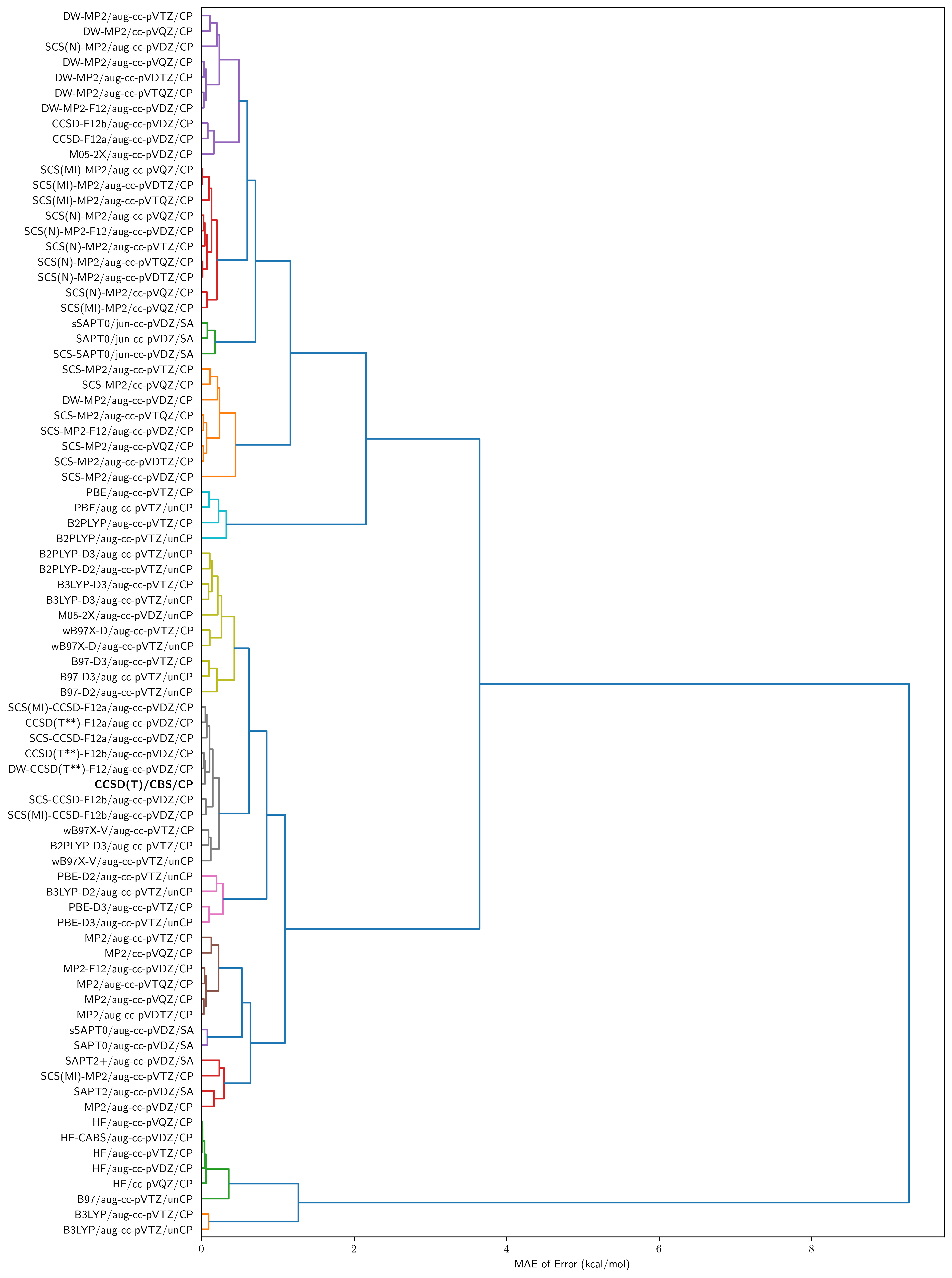}
    \caption{
        Dendogram of all-to-all $\delta$AP-Net2 model predicted error
        estimations ordered by MAE. Note the clusters of methods are
        nearly identical as the all-to-all M1 to M2 dendogram in the SI,
        meaning that the models are accurately predicting any M1 to M2.
    }
    \label{fg:dendogramSI}
\end{figure*}



\clearpage

\end{document}